\documentstyle[preprint,aps]{revtex}
\begin{document}

\title{
Pad\'{e} Interpolation: \\
Methodology and Application to Quarkonium}
\bigskip
\author{\bf C. N. Leung
and J. A. Murakowski\footnote{Present address:
Department of Electrical and Computer Engineering,
University of Delaware, Newark, DE 19716, U.S.A.}}
\address{Department of Physics and Astronomy,
University of Delaware\\
Newark, DE 19716, U.S.A. \\}
\maketitle
\bigskip

\begin{abstract}
A novel application of the Pad\'{e} approximation is proposed in
which the Pad\'{e} approximant is used as an interpolation for the
small and large coupling behaviors of a physical system, resulting
in a prediction of the behavior of the system at intermediate
couplings.  This method is applied to quarkonium systems and
reasonable values for the $c$ and $b$ quark masses are obtained.
\end{abstract}


\newpage
The Pad\'{e} approximation seeks to approximate the behavior of a
function, $f(x)$, by a ratio of two polynomials of $x$.  This ratio
is referred to as the Pad\'{e} approximant.  Compared to the usual
perturbative power series approximation, the Pad\'{e} approximant
has the advantage that it deviates less rapidly from the true
values of $f(x)$ as $x$ becomes large.  Recently, the Pad\'{e}
approximation has been applied to quantum field theories to
estimate the next order term in a perturbation series\cite{qft}.
The method involves calculating a certain physical quantity
perturbatively to $n$th order in the coupling constant and then
forming a Pad\'{e} approximant which, when expanded in a power
series of the coupling constant, reproduces the perturbative
result.  The $(n+1)$th order term in the expansion of the Pad\'{e}
approximant yields an estimate of the $(n+1)$th order term in the
perturbation series for the physical quantity.  It turns out one
can obtain reasonably good estimates from this approach.

In this paper a different usage of the Pad\'{e} approximation is
proposed.  We observe that, because of the nature of the Pad\'{e}
approximant, it can be expanded in a power series in $x$ when $x$
is small as well as in a power series in $\frac{1}{x}$ when $x$ is
large.  It is therefore interesting to ask the question: in cases
when both the small $x$ (e.g., weak coupling) and the large $x$
(e.g., strong coupling) behaviors of a theory can be computed
perturbatively, is it possible to form a single Pad\'{e}
approximant which interpolates the weak and strong coupling
behaviors, and if so, how well does this Pad\'{e} interpolation
approximate the behaviors of the theory at intermediate values of
the coupling constant?  This is a particularly timely question
since, with the advance of duality in supersymmetric gauge
theories\cite{dual}, we may someday be able to compute the strong
coupling behaviors of a theory from its dual theory.  The Pad\'{e}
interpolation will then provide a means to estimate the behaviors
of the theory for the entire range of the coupling constant.

The method proposed here goes beyond interpolating the strong
and weak coupling behaviors of a system.  For example, the
expansion parameter $x$ can be the temperature, the strength of
an applied field, or, as discussed below in the application to
heavy quarkonia, a parameter introduced to implement the Pad\'{e}
interpolation.

We have tested the Pad\'{e} interpolation method with examples
in which the exact result is known, with encouraging success.
To see how accurate the Pad\'{e} interpolation can be and to
illustrate the methodology involved, let us consider a simple
quantum mechanical two-state system with the Hamiltonian,
\begin{equation}
 H = \sigma_x + \lambda \sigma_z,
\label{ex}
\end{equation}
where the $\sigma$'s are the Pauli matrices and the coupling
constant $\lambda$ is assumed to be positive.  For $\lambda \ll 1$,
the $\sigma_z$ term may be treated as a perturbation and we find,
to second order in perturbation theory, the eigenvalues of $H$ are
\begin{equation}
 E^{<}_{\pm} = \pm 1 \pm \frac{\lambda^2}{2}.
\label{E}
\end{equation}
For $\lambda \gg 1$, the Hamiltonian can be written as $H =
\lambda (\sigma_z + \frac{1}{\lambda} \sigma_x)$, and the
$\sigma_x$ term can be treated as a perturbation.  We find, again
to second order in perturbation theory, the eigenvalues of $H$ are
now
\begin{equation}
 E^{>}_{\pm} = \pm \lambda \pm \frac{1}{2 \lambda}.
\label{E'}
\end{equation}

A Pad\'{e} approximant which interpolates the small and large
$\lambda$ behaviors of the energies can now be constructed.  For
the higher energy level $E_+$, we find
\begin{equation}
 E_{+}^{(\rm PA)} = \frac{\lambda^3 + \frac{3}{2} \lambda^2 +
        \frac{3}{2} \lambda + 1}{\lambda^2 + \frac{3}{2} \lambda
        + 1}.
\label{EPA}
\end{equation}
This Pad\'{e} approximant is uniquely determined from the
perturbative expansions for $E_+$ given in (\ref{E}) and
(\ref{E'}).  The large $\lambda$ behavior indicates that the
polynomial in the numerator must be one degree higher than the
polynomial in the denominator and that the coefficient for the
highest order term in $\lambda$ must be the same for the two
polynomials.  Without loss of generality, we may choose this
coefficient to be 1.  If the numerator is a polynomial of degree
$d$, there will be a total of $(2d - 1)$ coefficients to be
determined for the Pad\'{e} approximant.  Because the small
$\lambda$ behavior requires the numerator and the denominator to
have the same constant (i.e., $\lambda$-independent) term, there
are only $(2d - 2)$ remaining coefficients to be determined.
Expanding the Pad\'{e} approximant and matching against the
perturbation series in (\ref{E}) and (\ref{E'}) provide an
additional four conditions, which selects $d = 3$.

In Figure 1, the approximate result generated from Pad\'{e}
interpolation is compared with the exact result, $E_+ =
\sqrt{\lambda^2 + 1}$.  We see that the Pad\'{e} approximant
(open squares) tracks the exact result (solid curve) for all
values of the coupling constant.  In fact, $E_{+}^{(\rm PA)}$
differs from $E_+$ by no more than about 1\% for the entire
range of $\lambda$.  For example, for $\lambda$ = 0.5, 1.0, 2.0,
and 4.0, $E_{+}^{(\rm PA)}$ is larger than $E_+$ by 0.63\%,
1.02\%, 0.62\%, and 0.18\%, respectively.  One may improve the
approximation by calculating more terms in the perturbation
series for $E^{<}_+$ and $E^{>}_+$ and constructing the
corresponding Pad\'{e} approximant.  However, our example
suffices to demonstrate the potential power of the Pad\'{e}
interpolation method in that very few terms in the perturbation
expansions can yield a very accurate approximation to the exact
result for the entire range of the coupling constant.

For comparison, we have also plotted in Figure 1 the
perturbative result for small $\lambda$, $E^{<}_{+}$ (dotted
curve).  As expected, it only agrees with the exact result for
small values of $\lambda$ and diverges significantly from the
exact result when $\lambda$ becomes large.  Similarly,
$E^{>}_{+}$ will diverge from the exact result for small
$\lambda$.  In contrast, by interpolating $E^{<}_{+}$ and
$E^{>}_{+}$, the Pad\'{e} approximant is constrained not to
deviate too far from the exact result for the full range of the
coupling constant.  In this way, the Pad\'{e} interpolation
method can yield a very good approximation, provided the
quantity we try to approximate is a smooth, continuous function
of the coupling constant.

When applying the Pad\'{e} interpolation, one should beware of
potential unphysical singularities coming from the zeroes of
the polynomial in the denominator of the Pad\'{e} approximant.
This complication may limit the scope of applicability of the
method.  On the other hand, this property may prove useful in
some applications.  For instance, when interpolating the high
and low temperature behaviors of a system for which a phase
transition takes place at some intermediate temperature, one
may try to construct a Pad\'{e}-like approximant (perhaps
involving fractional powers in the polynomials) which mimics
the singular behavior near the phase transition point.

Another way to implement the Pad\'{e} interpolation method is
in cases when the Hamiltonian can be expressed as $H = H_1 +
H_2$, where the exact solutions for $H_1$ and $H_2$ (but not
$H$) are known.  Introducing the interpolating Hamiltonian,
\begin{equation}
 H(\beta) \equiv H_1 + \beta H_2,
\label{H}
\end{equation}
where $\beta$ is a positive constant, we can then treat $H_2$
as a perturbation when $\beta \ll 1$ and treat $H_1$ as a
perturbation when $\beta \gg 1$, in exactly the same way
as in the example (\ref{ex}).  A Pad\'{e} approximant is
formed interpolating the perturbative results for large and
small $\beta$.  Finally, an approximate solution for the
original Hamiltonian $H$ is obtained by setting $\beta$ equal
1 in the Pad\'{e} approximant.  This method will be applied
below to calculate quarkonium spectra.  Reasonable values for
the $c$ and $b$ quark masses are obtained by fitting the
calculated levels to their measured values, which demonstrates
the legitimacy of this Pad\'{e} interpolation approach.

Quarkonium refers to the bound state of a heavy quark $Q$ (e.g.,
$c$ or $b$ quark) with its antiquark $\bar{Q}$.  It is well
known that such systems can be described reasonably well using
nonrelativistic quantum mechanics\cite{QR}.  Various potential
energy functions have been used to model the $Q\bar{Q}$
interaction.  It has been found that the potential description
is flavor independent, i.e., the same potential describes
equally well the $c\bar{c}$ and the $b\bar{b}$ systems.  We
consider here a central potential consisting of an attractive
Coulomb term and a confining linear potential\cite{cl}:
\begin{equation}
 V(r) = - \frac{\alpha}{r} + \lambda r,
\label{V}
\end{equation}
where $\alpha$ and $\lambda$ are positive coupling constants.
We shall focus on the S states for the purpose of testing the
proposed interpolation method.  In this case, the Hamiltonian
for the radial Schr\"{o}dinger equation is simply
\begin{equation}
 H_r = - \frac{1}{2\mu} \frac{d^2}{dr^2} - \frac{\alpha}{r}
        + \lambda r,
\label{Hr}
\end{equation}
where $\mu$ is the reduced mass for the heavy quark $Q$,
$\mu = m_Q/2$.  Note that $H_r$ can be expressed as the sum of
two exactly solvable Hamiltonians: a Hamiltonian for the Coulomb
potential,
\begin{equation}
 H_C = - \frac{1}{4\mu} \frac{d^2}{dr^2} - \frac{\alpha}{r},
\label{Coulomb}
\end{equation}
and a Hamiltonian for the linear potential,
\begin{equation}
 H_L = - \frac{1}{4\mu} \frac{d^2}{dr^2} + \lambda r.
\label{linear}
\end{equation}
We have split the kinetic energy term in half so that the
``effective mass" that appears in $H_C$ and in $H_L$ is
$2\mu = m_Q$.

We may now form the interpolating Hamiltonian, $H_r(\beta) =
H_C + \beta H_L$, and perform perturbative calculations for
small $\beta$ as well as for large $\beta$.  We shall summarize
the results of our calculation here.  Details of the calculation
can be found in Ref.\cite{jam} where the calculation including
the quarkonium P states is also discussed.  (For the P states,
the centrifugal pontential energy term, $\frac{l(l+1)}
{2\mu r^2}$, should be included with $H_C$, resulting in a
solvable "hydrogen-like" Hamiltonian.  Because of the half
kinetic energy term, care must be taken to redefine the orbital
angular momentum quantum number in order to extract the energy
eigenvalues.)

The bound state energies for the S states of $H_r(\beta)$ are
computed for small $\beta$ as well as for large $\beta$ to the
same order in perturbation theory.  Wherever necessary (e.g.,
integrals involving the Airy functions, the eigenfunctions of
$H_L$), terms in the perturbation series are evaluated numerically.
In addition, for second and higher order calculations, the
infinite series that appear in the perturbation expansions are
estimated using the method of acceleration of convergence\cite{ac,jam}.
A separate Pad\'{e} approximant is formed interpolating the
small and large $\beta$ results from our first, second, and
third order calculations.  Our estimates for the S state
energies are obtained by letting $\beta$ equal 1 in the
respective Pad\'{e} approximant.  These are fitted to the
corresponding measured values treating $\alpha$, $\lambda$,
$m_c$, $m_b$ as well as the zero-point energies \(V_c\) (for
charmonium) and \(V_b\) (for bottomonium) as free parameters.
We used the data for the $J/\psi$(1S), $J/\psi$(2S),
$\Upsilon$(1S), $\Upsilon$(2S), and $\Upsilon$(3S) given in
Ref.\cite{RPP96}.  When performing the fit, care must be taken
to avoid the artificial singularities of the Pad\'{e} approximant.

The details of the fit results are presented in Table 1.  We
see that the first-order approximation already produces a
rather good fit to the measured S-state energies, although
the best fit values for $m_c$ and $m_b$ are somewhat high.
The second-order approximation improves the fit quality,
reproducing all of the quarkonium S-levels.  The fit quality,
defined as
\(
\sum_{i}\left(m_i^{\hbox{(experiment)}}-m_i^{\hbox{(calculated)}}
\right)^2
\),
worsens (from less than 1 to 240) as we go to the third-order 
approximation, primarily due to the increased difficulty to 
avoid a larger number of unphysical singularities from the 
Pad\'e approximant in performing the fit.  
Using the second-order results, our best fit values for $m_c$
and $m_b$ are 1.521 GeV and 5.046 GeV, respectively, to be
compared with the values given in Ref.\cite{RPP96}:
$m_c = 1.0 ~{\rm to}~ 1.6$ GeV; $m_b = 4.1 ~{\rm to}~ 4.5$ GeV.
The best fit values for the parameters in the quarkonium
potential (\ref{V}), $\alpha = 0.4984$ and $\lambda = 0.1771
~{\rm GeV}^2$, also compare favorably with earlier results:
$\alpha = 0.520$ and $\lambda = 0.183 ~{\rm GeV}^2$ in
Ref.\cite{EGKLY}; $\alpha = 0.507$ and $\lambda = 0.169
~{\rm GeV}^2$ in Ref.\cite{QR}.  

We have also performed a fit with the constraint \(V_b-V_c = 
2(m_b-m_c)\) on the model parameters.  The results are presented 
in Table 2. The fit quality in this case is comparable to that 
of the unconstrained fit and the best fit values of the model 
parameters differ somewhat from the best fit values of the 
unconstrained fit, indicating that the found minimum is not 
sharp and allows for some variation of the quark masses as long 
as their difference remains equal to half of the difference 
between the zero-point energies \(V_b\) and \(V_c\). The model 
parameters appear to be more stable than before as we go from 
first to second-order Pad\'e interpolation, which corroborates 
the physical significance of the constraint.  

As a final check, we have integrated numerically the 
Schr\"{o}dinger equation for the quarkonium systems with the 
second-order best fit values of the parameters to obtain the 
energy levels.  The results are shown in the last column of Table 1.  
This verifies the validity of the Pad\'{e} interpolation method.  

In conclusion, using quarkonia and a simple two-state model as
our testing grounds, we have shown that Pad\'{e} interpolation
can be a powerful method for estimating physical quantities at
intermediate values of the coupling constant where perturbative
calculations are not reliable.  There are many areas for which
this method may be applicable.  One of which is the K-meson
system.  The strange quark mass, $m_s$, has the value such that
neither chiral perturbation theory (for small quark masses) nor
heavy quark effective theory (for large quark masses) gives a
good description of K-meson properties.  With Pad\'{e}
interpolation we may be able to obtain a more accurate estimate
of the K-meson properties by interpolating the small $m_s$ and
large $m_s$ behaviors which can be obtained perturbatively
through chiral perturbation theory and heavy quark effective
theory, respectively.  These issues are being examined by one
of us and the results will be reported in the near future\cite{K}.

\vspace*{2.0 cm}
\centerline{\bf Acknowledgement}
\bigskip
This work was supported in part by the U.S. Department of Energy
under grant DE-FG02-84ER40163.

\newpage
\begin{center}
{\bf Table 1.} Fit results from Pad\'{e} interpolation.\\
\bigskip
\bigskip
\begin{tabular}{lccccc}
\hline \hline
Enrgy level & Measured & 1st order & 2nd order & 3rd order
& Numerical\\
\hline
\(J/\psi(1S)\) [MeV] & 3097  & 3097 & 3097 & 3089 & 3097\\
\(J/\psi(2S)\) [MeV] & 3686  & 3686 & 3686 & 3694 & 3687\\
\(\Upsilon(1S)\) [MeV] & 9460 & 9459 & 9460 & 9464 & 9456\\
\(\Upsilon(2S)\) [MeV] & 10023 &10026 & 10023 & 10028 & 10020\\
\(\Upsilon(3S)\) [MeV] & 10355 &10353 & 10355 & 10347 & 10356\\
Fit quality [MeV$^2$] & & 13 & $<1$ & 240 &\\
\hline \hline
Fit parameters &  &  &  &  & \\
\hline
\(\alpha\) & &0.4600 & 0.4984 & 0.7510 & 0.4984 \\
\(\lambda\) [GeV\(^2\)] & &0.1834 & 0.1771 & 0.1344 & 0.1771 \\
\(V_c\) [MeV] & &2767 & 2765 & 2953 & 2765 \\
\(V_b\) [MeV] & &9573 & 9585 & 9761 & 9585 \\
\(m_c\) [MeV] & &1719 & 1521 & 1253 & 1521\\
\(m_b\) [MeV] & &5538 & 5046 & 4143 & 5046 \\
\hline \hline
\end{tabular}
\end{center}

\newpage
\begin{center}
{\bf Table 2.} Fit results from Pad\'{e} interpolation
with the constraint \(V_b-V_c = 2(m_b-m_c)\).\\
\bigskip
\bigskip
\begin{tabular}{lccc}
\hline \hline
Enrgy level & Measured & 1st order & 2nd order \\
\hline
\(J/\psi(1S)\) [MeV] & 3097  & 3097 & 3098 \\
\(J/\psi(2S)\) [MeV] & 3686  & 3686 & 3685 \\
\(\Upsilon(1S)\) [MeV] & 9460 & 9459 & 9460 \\
\(\Upsilon(2S)\) [MeV] & 10023 &10026 & 10022 \\
\(\Upsilon(3S)\) [MeV] & 10355 &10353 & 10356 \\
Fit quality [MeV$^2$] & & 13 & 3 \\
\hline \hline
Fit parameters &  &  & \\
\hline
\(\alpha\) & &0.4850 & 0.4964  \\
\(\lambda\) [GeV\(^2\)] & &0.1741 & 0.1784 \\
\(V_c\) [MeV] & &2770 & 2773 \\
\(V_b\) [MeV] & &9571 & 9574 \\
\(m_c\) [MeV] & &1560 & 1572 \\
\(m_b\) [MeV] & &4960 & 4972 \\
\hline \hline
\end{tabular}
\end{center}

\newpage
\centerline{\bf FIGURE CAPTION}
\bigskip
\noindent{
{\bf Figure 1}: see description in the text.}

\end{document}